# SYSTEM SUPPORT FOR MANAGING INVALID BINDINGS

Lachhman Das Dhomeja[1], Yasir Arfat Malkani[1], Azhar Ali Shah[2]

and Khalil Khoumbati[2]

[1]School of Informatics, University of Sussex, Brighton, UK
`l.d.dhomeja@sussex.ac.uk` and `y.a.malkani@sussex.ac.uk`
[2]Institute of Information Technology, University of Sindh, Jamshoro, Pakistan
`azhar.shah@usindh.edu.pk` and `khalil.khoumbati@usindh.edu.pk`

## ABSTRACT

*Context-aware adaptation is a central aspect of pervasive computing applications, enabling them to adapt and perform tasks based on contextual information. One of the aspects of context-aware adaptation is reconfiguration in which bindings are created between application component and remote services in order to realize new behaviour in response to contextual information. Various research efforts provide reconfiguration support and allow the development of adaptive context-aware applications from high-level specifications, but don't consider failure conditions that might arise during execution of such applications, making bindings between application and remote services invalid. To this end, we propose and implement our design approach to reconfiguration to manage invalid bindings. The development and modification of adaptive context-aware applications is a complex task, and an issue of an invalidity of bindings further complicates development efforts. To reduce the development efforts, our approach provides an application-transparent solution where the issue of the invalidity of bindings is handled by our system, Policy-Based Contextual Reconfiguration and Adaptation (PCRA)[1], not by an application developer. In this paper, we present and describe our approach to managing invalid bindings and compare it with other approaches to this problem. We also provide performance evaluation of our approach.*

## 1. INTRODUCTION

In order for pervasive computing systems to be able to perform tasks which support us in everyday life without requiring attention from the users of the environment, they need to adapt themselves in response to context. This makes context-awareness in general and context-aware adaptation in particular, an essential requirement for pervasive computing systems. Context-awareness was pioneered by researchers at Xerox PARC Laboratory [2-4] under the vision of ubiquitous computing [5], also known as pervasive computing. Since then, this area has been widely studied and many context-aware systems have been built to demonstrate the usefulness of this research area. A summary of contributions made in this area can be found in [6,7].

Context-aware adaptation is a process of modifying the behaviour of the applications in response to change of context, which may be required to fulfil user needs, enrich the user experience, address inherent limitations of mobile technology (such as varying network quality, limited battery life, small screen size), etc.

---

[1] PCRA [1] is a policy-based context-aware adaptation system, which enables the development and execution of adaptive context-aware applications using Ponder2 [17,18] policy specifications.

Context-aware adaptation may involve restructuring or reconfiguring the software components of the applications to realize new behaviour in response to context. This can be achieved by discovering service(s) based on context and binding them to application components. For example, a user may want to have her messages displayed or printed to the nearest rendering device to her location. This requires discovering a device based on the location of the user and binding to it, and then sending messages to the bound device. As another example, application reconfiguration may be used to enrich the experience of a mobile user by providing her with a service of interest with respect to her changed location without requiring any cooperation from her. For instance, when the user is standing near a cinema, a movie information service could send information about the movies being exhibited in that cinema.

Context-aware adaptation may also involve modifying the application behaviour through the modification of the behaviour of the component/service in response to context. For example, in a simple home lighting scenario the light service may be modified to adjust the light value to some user-preferred value based on the user's activity. As another example, a video service may be modified to serve a text version of the content instead of audio / video content in response to a drop in network bandwidth.

Our primary research objective has been to provide a broader scope of adaptation and a high-level programming model to enable the development and execution of diverse adaptive context-aware applications, hence the design and development of PCRA. However, various failure conditions can arise during the execution of such applications, making bindings between application and bound services invalid and so require support to update them. The bindings may become invalid due to many reasons, such as sudden non-availability of the bound service (due to power failure at the hosting device where the bound service is running), or the bound service has been moved to some other location over the network for load-balancing purposes, or it has been moved closer to the entity accessing the bound service in order to save bandwidth.

Adaptive context-aware applications are complex to develop, maintain and modify, and the issue of managing these binding failures further complicates development efforts. In order to reduce the complexity involved in developing, maintaining and modifying adaptive context-aware applications, we propose to relieve an application developer from the responsibility for managing invalid bindings and to delegate this to our system, PCRA. To this end we propose and implement a design approach to reconfiguration to manage invalid bindings, which is integrated within PCRA. Other features of PCRA such as contextual reconfiguration support, contextual adaptation support and seamless caching support of virtual stub for improved performance are out of scope of this paper. In this paper we only focus on PCRA's reconfiguration support for handling the issue of invalidity of bindings.

The remainder of this paper is organised as follows. Section 2 provides background. Section 3 presents a high-level architecture of PCRA and briefly describes its parts. Section 4 presents and describes our approach to managing invalid bindings. Section 5 provides performance evaluation of our approach. Section 6 compares our approach to related work. Finally, section 7 concludes the paper.

## 2. BACKGROUND

An adaptive context-aware application may fail to function correctly due to failures in locating the required service during a reconfiguration process; or when the reconfiguration process successfully discovered a service and created a binding between the application and the found service based on context, an interaction between an application component and the bound service can be affected due to network-induced problems, such as time-outs, temporary network failure or power failure at the hosting device where the bound service is running. When these kinds of network-induced problems occur, remote exceptions are generated. These are traditionally handled by having an exception handler at the client that would try to invoke bound service a few times, and if it does not succeed it would attempt to discover a new service. The problem of power failure at the hosting device where the bound service is running is different because the bound service becomes unavailable, causing the binding to become invalid. The binding is said to be invalid when the reference (real proxy/stub) to the bound service becomes invalid. Even if the power of the hosting device is restored and the same service is run again, the reference to the bound service obtained before the power failure is no longer valid, causing all the bindings to this service to become invalid. Any interaction with the service through an invalid stub would result in a remote exception being generated.

There are other situations that may cause bindings to become invalid, for an instance, when the bound service is moved to another location over a network for load-balancing purposes, or it has been moved closer to an entity accessing that bound service in order to improve service provisioning or to save bandwidth. In these situations a reference to the bound service becomes invalid upon its migration to a new location, causing all the bindings between software components and the moved service to become invalid. The above discussion shows that there are two primary causes of invalidating a binding: one is due to power failure at the hosting device where the bound service is running and other is the migration of a bound service to a new location. The solution to this problem requires updating the invalid reference to the bound service. This problem can be solved by any of two design approaches: (1) the client itself must be responsible for handling this issue. When the bound service becomes unavailable or is migrated, the old reference to this service is no longer valid and if the client uses invalid reference to this service, an exception will be thrown. The client has to handle the exception, for example, updating the reference and repeating the call. (2) The system must be responsible for handling this issue where updating a reference is carried out by the system. In the former, the burden is on the client and such reconfiguration is not application transparent, while in the latter it is done by the system and such a reconfiguration is application transparent.

There are various research efforts focusing on this issue and provide system level approach to maintain a reference with a moving object, for an instance, [8, 9]. Other systems that provide application transparent support for managing references upon component migration or replacement include [10, 11]. There are other research efforts that provide high-level programming model for developing adaptive context-aware applications also consider failure conditions causing binding to become invalid, and these include [12, 13]. We also advocate and propose system level approach to managing invalid bindings and suggest a design approach, in which a system component, virtual stub (discussed later) is responsible for updating an invalid binding.

The concept of the virtual stub, also called smart proxy in the literature has been used as a design component to address various issues, and some of the research efforts that make use of this component include [11, 14, 15, 16]. The virtual stub/smart proxy wraps the real proxy of the remote service and provides more functionality than the real proxy does (forwarding remote calls from a client to remote service), depending on the requirements of the system. For example, it may be required to perform client-side validation before calling actual methods of a target object; it may be desirable to perform client-side caching to save the remote calls; it may be desirable that in case of any remote exception the client should not handle the exception, but instead smart proxy, so that the client is free to deal with real requirements of the applications, etc. Other additional responsibilities performed by the smart proxy may include performing security (e.g. not giving access to certain remote objects according to IP address) and load-balancing.

In our approach to updating invalid bindings, the virtual stub is also used as a design component which is responsible for performing reconfiguration to update bindings when bindings become invalid. Reconfiguration support to manage invalid bindings is integrated within PCRA and in order to be able to describe our approach, we first briefly present and describe a high-level architecture of PCRA.

## 3. HIGH-LEVEL DESIGN OF PCRA

PCRA is a policy-based context-aware adaptation system, which enables the development and execution of adaptive context-aware applications using Ponder2 policy specifications in which a policy is specified declaratively in an ECA (Event-Condition-Action) format. The overall architecture of PCRA is comprised of three parts as shown below in figure-1: the Ponder2 System, our Reconfiguration and Adaptation Infrastructure (RAI) and Java RMI.

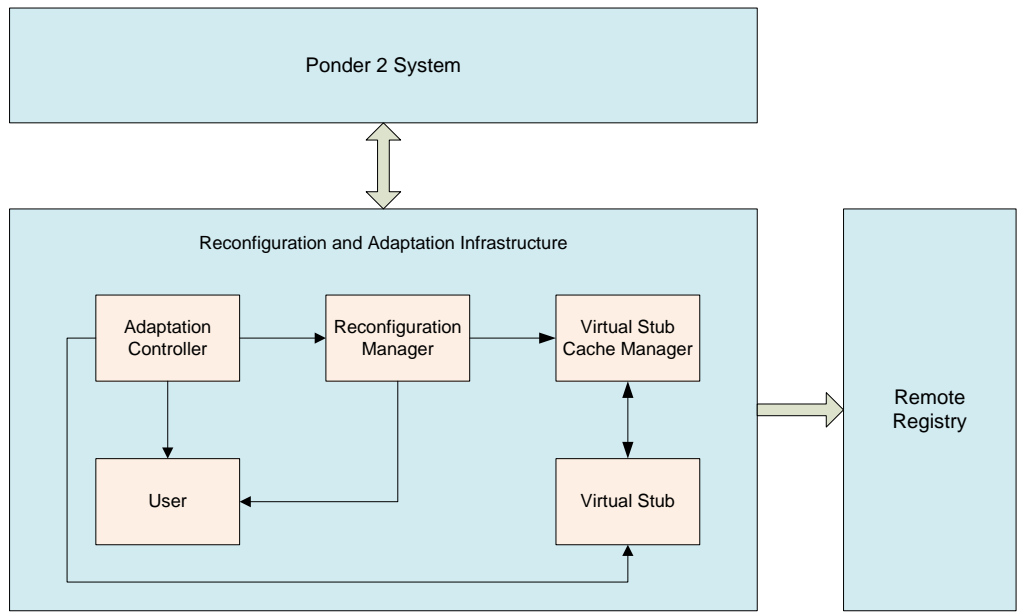

Figure-1: High-level System Architecture of PCRA

### 3.1 Ponder2 System

Ponder2 [17, 18] is a re-design and re-implementation of Ponder [19], which is used by many in both academia and industry. Ponder2 is a light-weight, self-contained, extensible and scalable policy system for specifying and enforcing policies, which can be used at different levels of scale from small resource-constrained devices (e.g., PDA and mobile phones) to complex environments. Ponder2 provides the support for both obligation policies and authorization policies, and has Events, Policies, an Obligation Policy Interpreter, Authorization Policy Interpreter, Command Interpreter and Domain Service. Within PCRA, adaptive context-aware applications are developed by expressing adaptive behaviour of the applications using Ponder2 obligation policies.

### 3.2 Reconfiguration and Adaptation Infrastructure (RAI)

Our main contribution within PCRA is the design and development of RAI, which provides reconfiguration support for managing invalid bindings in addition to other features such as runtime support for contextual reconfiguration, contextual adaptation and seamless caching support of virtual stubs for improved performance. This paper only focuses on RAI's reconfiguration support for managing invalid bindings, and description of other features of this infrastructure are out of the scope of this paper. The core design component of RAI which implements the support for managing invalid bindings is the virtual stub. However, in order to be able to demonstrate our approach to reconfiguration to recover from binding failures, a brief description of all RAI design components is given below.

### 3.2.1 Virtual Stub

Virtual stub holds the real stub of the remote service and the binding to that service is created through its corresponding virtual stub. We shall see later how the bindings are created. An application component communicates with a bound service not directly by using a real stub of the bound service, but by using a virtual stub. Virtual stub receives remote calls from the clients and forwards them to the remote service that it is representing. In addition to forwarding remote calls, when the binding to the remote service which it represents becomes invalid for any reasons (for whatever reason), the virtual stub performs reconfiguration to update invalid binding. We describe our approach in section 4.

### 3.2.2 User Component

This component models the users of the environment and holds bindings to remote services. Among other functionality, this component also saves the user preferences for various services and includes code through which the user can set or customize her preferences dynamically. We describe later how bindings are created between the user instance and the remote services based on context.

### 3.2.3 Reconfiguration Manager

This component is responsible for creating bindings to remote services based on context. When a context event occurs, the appropriate policy would be triggered. The policy would interact with the reconfiguration manager to perform reconfiguration, creating bindings between the user instance and remote services. The other component that reconfiguration manager interacts with in creating bindings is the Virtual Stub Cache Manager, which is discussed next.

### 3.2.4 Virtual Stub Cache Manager

This component implements the functionality for seamless caching support of virtual stubs for improved performance, one of the features of RAI. In addition to this, it works hand-in-hand with reconfiguration manager in process of creating bindings to remote services. During the binding process, the reconfiguration manager communicates with the virtual stub cache manager and asks for virtual stubs for each of remote services. In response to this, the virtual stub performs lookup to discover the real proxy for each of remote services, creates an instance of a virtual stub for each of the services and initializes it with the corresponding real proxy, caches all virtual stubs and then hand them to reconfiguration manager. When the reconfiguration manager has received the virtual stubs, it hands them to the user instance.

Once the user instance has the virtual stubs, this means that the bindings have been created between the user instance and the remote services. This completes the creation of bindings between the user of the environment and the remote services based on context. From above binding process, it can be noted that the virtual stub wraps the real proxy of the remote service, and the binding to that service is through its corresponding virtual stub.

### 3.2.5 Adaptation Controller

This design component provides the implementation of contextual adaptation support, one of the features provided by RAI. This component responds to adaptation policies and performs adaptation in terms of modifying the behaviour of the service. As the binding to the service is represented by its corresponding virtual stub, the adaptation controller communicates with the service through its virtual stub and adapts the behaviour of the bound service through parameter adjustments.

### 3.3 RMI

RMI is a distributed middleware that provides the means for developing remote services and remote method communication between these services and their clients. Within PCRA, remote services are developed using RMI and registered in RMI registry. As a part of binding process, these services are discovered and bound to based on context. To adapt the behaviour of these services, their methods are invoked using RMI protocol.

All three parts of the overall architecture work together to provide the system, PCRA, in which adaptive context-aware applications can be developed using polices. As can be noted from above discussion, PCRA architecture provides infrastructure for specifying and enforcing policies via Ponder2 system, and also provides a broader support of adaptation (contextual reconfiguration, contextual adaptation) and the support for reconfiguration to recover from invalid bindings through our RAI. All the developer or end user is required to do for developing adaptive context-aware application is to express binding policies and adaptation policies in Ponder2 specifications.

## 4. OUR APPROACH TO MANAGING INVALID BINDINGS

In section 3.2 we briefly described all design components of our RAI, including the virtual stub. The key to our approach is the use of virtual stub as a design component, which implements the support for reconfiguration to recover from invalid bindings. Virtual stub in RAI as discussed in section 3.2.1, in addition to forwarding remote calls to the remote service, has an additional responsibility of performing reconfiguration to recover from binding failures. As discussed in background section, various situations might arise during the execution of adaptive context-aware applications, causing bindings to remote services to become invalid. In all these situations, the real proxy of the bound service becomes invalid, causing all the bindings to this service to become invalid.

Figure-2 demonstrates our approach and figure-3 shows the partial code of virtual stub. In our approach to reconfiguration for managing bindings, when the binding becomes invalid due to any of these reasons and a method call is made on an invalid reference, an exception is thrown and caught by the virtual stub. In response to the exception, the virtual stub immediately performs the reconfiguration to update the invalid reference and repeats the call.

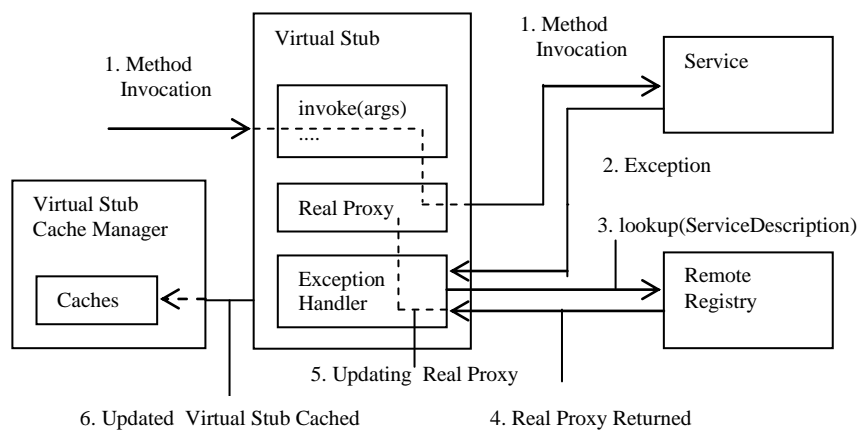

Figure-2: Reconfiguration to manage bindings

```
1. public class VirtualStub implements  Serializable {
2.
3.     Remote r = null;
4.     String serviceItholds;
5.     …………
6.     …………
7.     …………
8.
9.     void invokeMethod(String methodName, String arg){
10.
11.      Class[] partypes;
12.      remoteref= r.getClass();
13.      if(Test.isInteger(arg)){
14.          partypes=new Class[]{Integer.TYPE};
15.          parameter=(Object)Integer.valueOf(arg);
16.      }else{
17.          partypes=new Class[]{String.class};
18.          parameter = (Object)arg;
19.      }
20.
21.      try {
22.
23.          Method method = remoteref.getMethod(methodName,partypes);
24.          method.invoke(r,parameter);
25.      } catch (InvocationTargetException e) {
26.          invalidReference();
27.           this.invokeMethod(methodName, arg);
28.      }
29.      …………
30.      …………
31.      …………
32.    }
33.
34.    void invalidReference(){
35.     // does lookup, updates invalid reference, and caches it.
36.     try {
37.       r=Naming.lookup(serviceItholds);
38.       VirtualStubCache.virtualStub_Table.put(serviceItholds,
39.       VirtualStubCache.serializeVirtualStubIntoString(this));
40.
41.     } catch (RemoteException e1) {
42.
43.     }
44.     …………
45.     …………
46.     …………
47.    }
48. }
```

Figure-3: Partial source code of virtual stub

As a result of a remote method call (figure 3, line 24) on the invalid proxy, the exception is thrown, which is caught by virtual stub (figure 3, line 25). As a response, the virtual stub performs the following actions.

- It invokes its invalidreference() (figure 3, line 26), which performs a remote lookup (figure 3, line 37) to obtain a new copy of the real proxy and the invalid proxy is replaced by new one, thus updating the invalid reference.
- The virtual stub Cache Manager still has the virtual stub saved which contains invalid proxy. The virtual stub communicates with the virtual stub cache manager and sends a copy of itself (figure 3, line 39), which now contains the updated proxy. The virtual stub cache manager replaces the old copy with new one into cache.

- It then repeats the remote call (figure 3, line 27).

The message sequence diagram below (figure-4) captures the sequence of messages involved in reconfiguration to recover from binding failure.

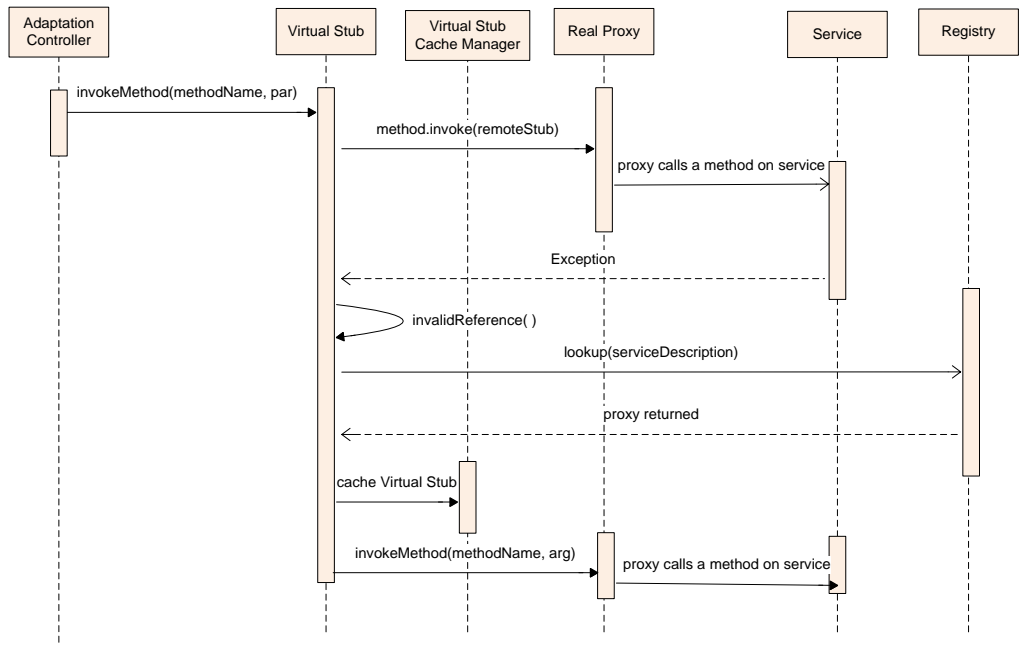

Figure-4: Message sequence diagram for reconfiguration to manage invalid bindings

## 5. PERFORMANCE EVALUATION

To study the cost of our approach to managing invalid bindings, we have conducted two tests. In test 1 we measure the reconfiguration time to manage invalid binding under a local setting, while in test2, the reconfiguration time is measured in a distributed setting. Reconfiguration time is a time taken by PCRA to update the invalid binding and then to repeat the call on the bound service. This time is measured from the point when exception is received by the virtual stub in response to a call on an invalid proxy of remote service held by the virtual stub until a new copy of the proxy is obtained and the call is repeated on the bound service. Figure-5 shows sequence diagram for reconfiguration time.

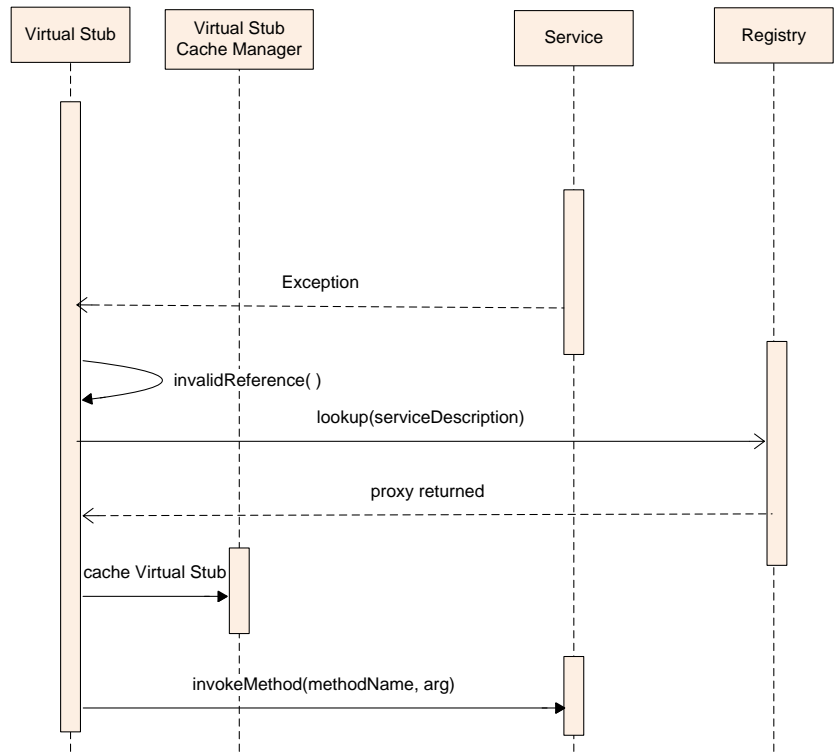

Figure-5: Sequence diagram of reconfiguration time

As can be noted from above figure, reconfiguration time involve a lookup time, time taken to cache an updated copy of virtual stub and the time taken by a repeated remote method call.

### 5.1 Test environment

We conducted tests under both a local setting and a distributed setting. For these tests we used two machines of the same specifications - Intel Pentium (4) 1.8 GHz and 256 RAM. Both machines were running Windows XP and used JDK 1.5. The JDK 1.5 or above is the requirement for running PCRA. In the local setting, PCRA (Ponder2 system and Reconfiguration and Adaptation Infrastructure (RAI)), RMI registry, remote services were running on the same machine.

In a distributed setting, two machines of the same specifications (mentioned above) were connected through a 100-Mbit Ethernet. PCRA (Ponder2 and reconfiguration and adaptation infrastructure) were running on one machine, while a RMI registry and remote services were running on other machine.

### 5.2 Test Results

Below we present the results obtained from the tests under the local setting and distributed setting.

#### 5.2.1 Local Setting

**Test 1:** Reconfiguration time in local setting

In local setting, PCRA (Ponder2 and RAI), RMI registry and remote service all were running on the same machine in three separate processes. As discussed in section 2, one of the reasons for the binding to become invalid is a power failure on the host machine where the bound service is running. After the power of the hosting device is restored and the bound service is run again, the proxy of the bound service obtained before the power failure is no longer valid, resulting in all the bindings to this service to become invalid. We simulated this power failure situation by closing the process under which remote services were running and then restarting. We ran a follow me service hypothetical example scenario (one of the example scenarios we have tested on PCRA, and for detailed description and its source code see section 4.1.2.2 in [1]) on PCRA in which a binding to a printer service or screen service is created based on context (location of the user). After the binding to printer service / screen service was created, we closed the windows where printer service and screen service were running and then restarted. This invalidated the binding, hence any call on the printer service / screen service resulted in an exception being thrown. We measured reconfiguration time from the point exception was received until binding was updated (by discovering the same service again and updating the invalid proxy) and then the call on the bound service was repeated. The test case was performed 20 times.

**Test 2:** Reconfiguration time in distributed setting

In distributed setting, PCRA (Ponder2 and RAI) was running on one machine, while RMI registry and remote services were running on other machine where both were running in separate processes. To measure the reconfiguration time the process running the remote services was closed and then restarted, and this process was repeated 20 times.

**Results:** The reported reconfiguration time for test 1 is an average time in local setting, while reported reconfiguration time for test 2 is an average time in a distributed setting. Both times are presented graphically in figure-6.

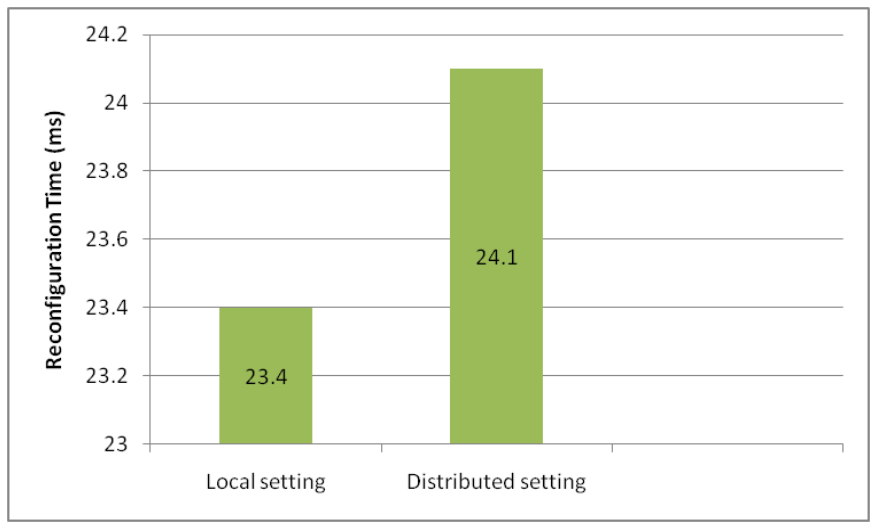

Figure-6: Reconfiguration time in local setting and distributed setting

As discussed before, reconfiguration time involves lookup time, time taken to cache an updated copy of virtual stub and a repeated remote call. This means a total reconfiguration time taken by PCRA to manage invalid bindings is a sum of these three times. The two times, lookup time and a repeated method call provide largest contributions to reconfiguration time as these are remote calls and the remote call is at least 1000 times slower than a local call.

## 6. COMPARATIVE ANALYSIS

In this section, we provide comparative analysis of our approach to other approaches to the same problem. In [12], the authors identify various failure conditions and robustness issues that can arise in context-aware pervasive computing applications and provide application-level recovery mechanisms. However, we only focus on binding failures that affect an interaction between an application and the bound service. The binding failures can be caused by either power-failure problems, or migration or replacement of the bound service. Although they handle various failure conditions and hence their solution is comprehensive, their solution is not application transparent. In contrast, our reconfiguration approach to managing bindings is application transparent.

Most of the current mobile frameworks use a system level approach to maintain a reference with a moving target object and they normally solve this problem with one of the two approaches. The first approach is to continuously maintain a valid reference to the moving target using a tracking mechanism as done in [8,9]. The tracker is a forwarding pointer. Upon the migration of a component to other location, the system creates a tracker in the old place. The method calls are forwarded to the moved target object by that tracker. While this approach provides an application transparent mechanism to always maintain a valid reference to the moved component, it is costly in two respects: (1) the system must create a tracker in the old place and (2) the method calls are first received by the tracker and then forwarded to the moved object. The second approach is to rebind the reference to the relocated object each time a method of the target object is called so that the reference to the moved target is always valid. This approach is very costly from the system performance point of view because each time the target method is called, a lookup operation is performed and this is very time consuming. In contrast, our approach does not suffer from these system performance issues.

Other systems that provide application transparent support for managing references upon component migration or replacement include [10,11]. These systems use a system design component called a virtual stub/smart proxy that holds/wraps the real proxy of the target component. An interaction between the application component and target object takes place through the virtual stub/smart proxy. In an attempt to call a method on an invalid reference, the virtual stub/smart proxy performs reconfiguration to update the invalid reference, thus providing an application transparent reconfiguration. In [10], the smart proxy catches an exception generated by an attempt to invoke a method on a target object which has since been replaced, and attempts to update the reference by looking up the replacement object in a naming service. In [11], a virtual stub updates the invalid reference when asked by the system or in response to an exception, generated by an attempt to invoke a method call on the target object which has since been moved or replaced. Our approach is similar to theirs in that our approach also uses a virtual stub and thus it is application transparent, but it is different in that the virtual stub reflectively invokes remote methods of the bound service. This allows having a generic virtual stub definition which can be used to wrap a real proxy to any remote service without knowing a remote interface implemented by the remote service. This means the same virtual stub definition is used for each unique remote service without the need for having the definition of each remote interface implemented by a remote service available locally. This results in a reduction in total amount of code.

## 7. CONCLUSION

In this paper we have presented and described our reconfiguration approach to managing invalid bindings. The key design component of our approach was the virtual stub, which implemented most of the functionality of our approach. The binding to any service involved discovering a remote service and obtaining its real proxy, creating an instance of virtual stub and initializing it with the obtained real proxy and then handing virtual stub to the user instance. Any method call to bound service through its corresponding virtual stub with an invalid proxy resulted in an exception thrown which was received by the virtual stub. In response to this exception, the virtual stub updated invalid binding (by discovering a new copy of real proxy and updating invalid proxy) and then repeated the call on the bound service. As invalid bindings are managed by the virtual stub, which is a system component, not by the application developer, our approach is an application transparent. As a result it contributes to reducing the development efforts for developing adaptive context-aware applications.

**Authors**

Dr. Lachhman Das Dhomeja is an Assistant Professor at the Institute of Information & Communication Technology (IICT), University of Sindh, Jamshoro, Pakistan. He got his Master's degree in Computer Technology from University of Sindh, Jamshoro (Pakistan) in 1991 and PhD from University of Sussex, UK in 2011. His main research area is Pervasive Computing in general and policy-based context-awareness in particular. His other research interests include secure device pairing in ubiquitous environments and Distributed Computing. 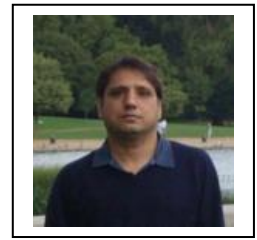

Dr. Yasir Arfat Malkani is a Lecturer at the Institute of Mathematics and Computer Science (IMCS), University of Sindh, Jamshoro, Pakistan. He go his Master's degree in Computer Science from University of Sindh, Jamshoro (Pakistan) in 2003 and PhD from University of Sussex, Brighton, UK in 2011. His main area of research is Pervasive Computing. His research is focused on secure device/service discovery and access control mechanisms using policies and location/proximity data/information. He is also interested in sensor networks, wireless networks (including WiFi, Bluetooth, WiMAX, etc), and solutions to various issues in distributed and pervasive computing systems through the integration of tools and techniques from distinct disciplines/areas. 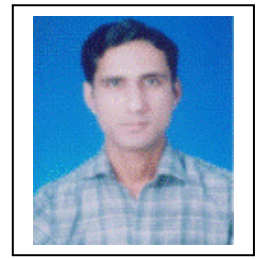

Dr. Azhar Ali Shah received the M.Sc. degree in electronics and the M.Phil. degree in information technology from the University of Sindh, Jamshoro, Pakistan, in 1998 and 2004, respectively. He recieved Ph.D. degree in Computer Science from University of Nottingham, Nottingham, U.K in 2011. He is currently an Assistant Professor in the Institute of Information and Communication Technologies, University of Sindh. His research interests include the use of grid-styled parallel/distributed computing for grand challenge applications (GCAs) in the field of life sciences, new trends in software engineering, pervasive computing, localization and internationalization in the context of regional languages of Pakistan. 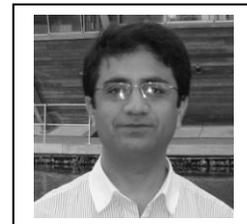

Dr. Khalil Khoumabti is a Professor in the Institute of Information and Communication Technology University of Sindh, Jamshoro, Pakistan. He gets hold of his PhD from the School of Information Systems, Computing and Mathematics, Brunel University, UK, M.Sc in Computer Technology and B.Sc in Electronics from Institute of Information Technology University of Sindh, Jamshoro, Pakistan. His current research focus is on the adoption of enterprise application integration in healthcare organization and e-government applications. He is on the editorial board of several international journals such as Transforming Government: People, Process and Policy. He has published 26 research papers in internationally refereed journals such as Journal of Management Information Systems and Journal of Computer and Information Systems. He has presented 20 papers in international conferences. 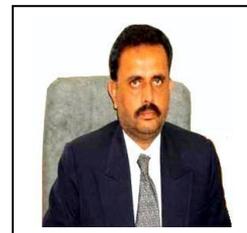